\documentclass[journal,twoside,web]{ieeecolor}
\usepackage{tmi}
\usepackage{cite}
\usepackage{amsmath,amssymb,amsfonts}
\usepackage{graphicx}
\usepackage{textcomp}
\usepackage{booktabs}
\usepackage{array}
\usepackage[ruled,vlined,linesnumbered]{algorithm2e}
\usepackage[final]{microtype}
\usepackage{threeparttable}
\usepackage{multirow}
\usepackage{url}
\usepackage{hyperref}
\def\BibTeX{{\rm B\kern-.05em{\sc i\kern-.025em b}\kern-.08em
    T\kern-.1667em\lower.7ex\hbox{E}\kern-.125emX}}
\title{3D-GlioPREDICT: 3D Latent Diffusion for Post-Radiotherapy Brain MRI Prediction in Patients with Glioma}
\author{Nordin Belkacemi, Selena Huisman, Vera C. Keil, Joost J.C. Verhoeff, and Szabolcs David, \IEEEmembership{Member, IEEE}
\thanks{Manuscript received XXXX XX, 2026; revised XXXX XX, 2026.}
\thanks{N. Belkacemi, S. Huisman, J. J. C. Verhoeff, and S. David are with the Department of Radiation Oncology, Amsterdam UMC, De Boelelaan 1117, 1081 HV Amsterdam, The Netherlands (e-mail: n.belkacemi@amsterdamumc.nl; s.i.huisman@amsterdamumc.nl; joost.verhoeff@amsterdamumc.nl; s.david@amsterdamumc.nl).}
\thanks{V. C. Keil is with the Department of Radiology and Nuclear Medicine, Amsterdam UMC, De Boelelaan 1117, 1081 HV Amsterdam, The Netherlands (e-mail: v.c.w.keil@amsterdamumc.nl).}}
\begin{document}
\maketitle

\begin{abstract}
Radiotherapy is a cornerstone of glioma treatment inducing complex structural changes in brain tissue that are difficult to anticipate. Predicting these changes from pretreatment data could improve understanding of treatment-related effects and support the development of image-based outcome prediction methods. Recent studies have shown that follow-up brain magnetic resonance imaging can be synthesized from baseline imaging and treatment information, but most existing approaches operate on single 2D slices and represent treatment as a global parameter, rather than a spatially dynamic variable. In this work, we address both limitations with a 3D latent diffusion framework that conditions image generation on the spatially resolved voxel-wise dose distribution, alongside a pretreatment image and follow-up time. To make volumetric synthesis computationally feasible, the model combines latent-space compression with ControlNet-based spatial conditioning. The method was trained and evaluated on a public dataset comprising 257 scans from 25 glioma patients. Prediction quality was assessed using mean squared error, peak signal-to-noise ratio, and structural similarity index. Anatomical consistency was further evaluated using Dice scores for cerebrospinal fluid, gray matter, and white matter segmentations, together with hippocampus volume prediction error and deformation analysis based on log Jacobian determinant maps. Compared with our previously proposed 2D approach, the 3D model achieved improved image similarity while maintaining good agreement with ground truth anatomy and deformation patterns. Overall, these results support the feasibility of 3D treatment-aware generative modeling for predicting post-radiotherapy brain MRI using only pretreatment information. Code is available at \url{https://github.com/nordinbelkacemi/fu-pred-3d}.
\end{abstract}


\section{Introduction}
\IEEEPARstart{G}{lioblastoma}, IDH-wildtype, is one of the most aggressive cancers with a median survival of approximately 14 months \cite{stuppRadiotherapyConcomitantAdjuvant2005} and the highest years of life lost ($>$20 years) \cite{burnetYearsLifeLost2005} despite maximal treatment consisting of surgery and chemoradiation. While this regimen represents the standard of care, it inevitably causes collateral damage to the surrounding brain, leading to progressive and irreversible cognitive decline in 50 to 90\% \cite{makaleMechanismsRadiotherapyassociatedCognitive2017, vankesselTumorrelatedNeurocognitiveDysfunction2017} of the patients. Ultimately, both the tumor and its treatment contribute to these deficits through different but converging mechanisms. For clinicians, treatment planning therefore remains a difficult trade-off between tumor control and preservation of neurological functioning and quality of life (QoL). Yet this trade-off cannot currently be personalized, because of the lack of tools to predict how an individual patient’s brain will respond to the standardized chemoradiotherapy plan. 

Medical imaging plays a crucial role in managing this delicate balance. Physicians establish baseline measurements using pretreatment Magnetic Resonance Imaging (MRI) and monitor both disease progression and treatment-related brain changes through longitudinal data acquired at regular intervals. Analyzing this data across patient groups can reveal patterns and consistent changes in brain tissue after treatment. In our previous work, we reported dose-dependent morphological changes in healthy brain tissue after radiotherapy in patients with glioblastoma, including localized white and gray matter volume loss \cite{huismanQuantifyingPostradiationAccelerated2022}, cortical thinning in healthy cortex \cite{nagtegaalEffectRadiationTherapy2020}, and volume loss in subcortical deep gray matter structures \cite{nagtegaalDosedependentVolumeLoss2021}. Such systematic patterns of post-treatment change may be learned by deep neural networks. Generative models are particularly well-suited for this task as they can model complex image distributions and synthesize future scans conditioned on baseline imaging and treatment information. The ability to predict post-treatment brain tissue changes before the start of treatment could support the evaluation of treatment plans and improve understanding of dose-related tissue response in vulnerable brain regions.

Longitudinal image prediction has been explored for several years in neuroimaging, particularly in the context of aging-related tissue change and the progression of neurodegenerative disease \cite{raviDegenerativeAdversarialNeuroimage2019, raviDegenerativeAdversarialNeuroimage2022}. By contrast, explicitly treatment-aware prediction remains relatively uncommon, especially for patients with intracranial tumors. Recent work has begun to address this by incorporating treatment information into conditional image generation frameworks \cite{liuTreatmentawareDiffusionProbabilistic2025, huismanRectifiedFlowbasedPrediction2026}. However, many of these approaches remain based on 2D slices, which limits the amount of anatomical context available to the model and discards through-plane information. This is not only a practical limitation, but also a conceptual one: the brain is a highly interconnected organ, and focal treatment-related injury may lead to distributed effects through structural disconnection and interactions between local and distant regions \cite{thiebautdeschottenEmergentPropertiesConnected2022}. Slice-wise models therefore have limited capacity to represent how disease- or treatment-induced changes propagate across the three-dimensional anatomy and connected tissue systems. From a clinical perspective, this is also misaligned with standard radiological practice, in which assessment is performed on full 3D volumes rather than isolated 2D slices.

While 3D generative techniques offer a principled way to overcome the limitations of slice-wise models, they also introduce substantial computational challenges due to the memory demands of volumetric synthesis. In this work, we develop a 3D latent diffusion framework to predict post-radiotherapy brain MRI scans in patients with glioblastoma. We adapt Medical AI for Synthetic Imaging (MAISI), a state-of-the-art volumetric latent diffusion architecture \cite{guoMAISIMedicalAI2024}. Specifically, we combine latent-space compression with ControlNet-based conditioning to generate follow-up MRIs from a pretreatment baseline MRI, a spatially registered radiotherapy dose map, and the target time after treatment. This formulation enables a prediction from treatment-planning information alone while explicitly modeling the spatial distribution of radiation in brain anatomy. We assess the anatomical fidelity of the predicted changes through complementary measures of image similarity, tissue-level anatomical agreement, brain-region volumetry, and deformation-based morphometry. This combined evaluation provides a more comprehensive view of model performance, particularly for subtle anatomical changes that may not be adequately captured by conventional image similarity metrics alone. Although imaging captures only part of the clinical picture, anatomically faithful predictions may still provide meaningful correlates of treatment response, linking observable structural change to clinically relevant outcomes that are not directly encoded in the images themselves.

\section{Related Work}
\label{sec:related}

\subsection{Longitudinal Image Prediction}
Early longitudinal brain MRI prediction methods were primarily formulated to capture endogenous neurobiological trajectories, particularly healthy aging and neurodegenerative progression. Within this line of work, Ravi et al. introduced DANI-Net in 2019 as an adversarial framework for subject-specific simulation of disease evolution in structural brain MRI \cite{raviDegenerativeAdversarialNeuroimage2019}, and subsequently extended this formulation in 2022 with 4D-DANI-Net, which incorporated spatiotemporal regularization to enforce anatomically plausible and temporally consistent longitudinal trajectories across multiple timepoints \cite{raviDegenerativeAdversarialNeuroimage2022}. These approaches therefore model intrinsic disease-related evolution rather than external interventions, and do not explicitly account for treatment-induced changes such as those associated with radiotherapy.

A major step toward treatment-conditioned longitudinal MRI generation was made by Liu et al. with Treatment-aware Diffusion (TaDiff) \cite{liuTreatmentawareDiffusionProbabilistic2025}. TaDiff conditions future MRI synthesis and tumor growth prediction on sequential prior MRI together with treatment and time information, using a joint diffusion and segmentation framework. Relative to earlier longitudinal prediction methods, this extends the conditioning space beyond imaging alone, allowing treatment history to influence the generated trajectory.

However, the treatment representation in TaDiff is not spatially resolved. Treatment is encoded as paired treatment-type and treatment-day variables embedded as global conditioning vectors, rather than as voxel-wise treatment descriptors, for example, chemoradiation at day 36 and temozolomide at days 64 and 127. This is a limitation for radiotherapy-oriented prediction, since radiation response depends strongly on the local dose distribution across brain tissue. Moreover, TaDiff requires multiple prior longitudinal MRI scans at inference time, making it less applicable to pretreatment prediction settings in which only baseline imaging and the planned treatment information are available.

In our recent work, Huisman et al. \cite{huismanRectifiedFlowbasedPrediction2026} addressed several of these limitations in the context of post-radiotherapy MRI prediction, using the same public SAILOR dataset as TaDiff for its experiments. Specifically, a follow-up brain MRI is generated from pre-radiotherapy baseline MRI, spatially registered radiotherapy dose maps, target follow-up time, and chemotherapy information using a 2D rectified flow framework. Compared with TaDiff, this formulation is better aligned with radiotherapy applications in two respects. First, the radiotherapy dose map is incorporated as an explicit spatial conditioning signal, allowing treatment to be represented as a localized prescription rather than only through global treatment labels. Second, inference relies on pretreatment inputs alone, making the framework applicable to prospective and counterfactual prediction settings in which post-treatment MRI is not yet available. In addition, the use of rectified flow reduces the number of sampling steps required for image generation relative to conventional DDPM-based approaches.

At the same time, the method of Huisman et al. remains slice-based. Predictions are generated on individual axial 2D slices rather than on full 3D volumes, which limits the model's ability to capture through-plane anatomical continuity and volumetric dose-response relationships. This is a relevant limitation in brain imaging, where both anatomy and radiation effects are inherently three-dimensional. Current work builds directly on this formulation by retaining the clinically relevant conditioning variables of pretreatment MRI, radiotherapy dose, and follow-up time, while extending the generative framework to full 3D latent diffusion.

\subsection{3D Generative Models in Medical Imaging}
Generating high-resolution 3D medical volumes is computationally demanding because memory requirements scale with the total number of voxels and therefore approximately cubically with linear image resolution \cite{wangDiffusionModels3D2025}. Latent Diffusion Models (LDMs) \cite{rombachHighResolutionImageSynthesis2022} address this by first compressing images into a lower-dimensional latent space with a variational autoencoder, then performing the diffusion process in that latent representation rather than in voxel space. Rombach et al.~\cite{rombachHighResolutionImageSynthesis2022} introduced this framework for high-resolution 2D image synthesis, while Pinaya et al.~\cite{pinayaBrainImagingGeneration2022} adapted it to 3D brain MRI, showing that latent diffusion can produce realistic neuroanatomy at a fraction of the memory cost of voxel-space diffusion. While these early models perform well, they were not formulated for treatment-aware longitudinal prediction with heterogeneous conditioning inputs such as baseline MRI and spatial radiation dose distributions. However, temporal and treatment-related conditioning can be incorporated with minimal architectural modification, as shown in subsequent work \cite{liuTreatmentawareDiffusionProbabilistic2025}, in which time points and treatment identifiers are each embedded as scalars via a learned embedding layer and a single-layer MLP.

To handle conditioning signals with spatial structure, a different mechanism is needed. One common approach is ControlNet \cite{controlnet}, originally proposed for 2D image synthesis tasks such as pose-guided generation, depth-conditioned synthesis, and edge-to-image translation. Rather than conditioning only through low-dimensional embeddings, ControlNet augments a pretrained diffusion U-Net with a trainable control branch that mirrors the downsampling and middle blocks of the locked backbone. The spatial input is encoded through this parallel pathway, and the resulting multi-scale control features are injected into the backbone through zero-initialized convolutions at the corresponding resolutions. In this way, the model can preserve the pretrained generative prior while gaining sensitivity to structured spatial guidance, enabling localized and anatomically specific control over the synthesis process.

Recent work has advanced 3D generation further, introducing multi-stage pipelines that offer latent diffusion, while also supporting downstream tasks through ControlNet conditioning. Wang et al.~\cite{wang3DMedDiffusion3D2024} introduced 3D MedDiffusion, which combines a patch-volume autoencoder with a dual-flow diffusion backbone called BiFlowNet. Their autoencoder encodes small 3D volume patches of the input image, quantizes them with a learned vector quantized codebook into compact latent tokens, and then decodes the assembled latent volume jointly to reduce patch boundary artifacts. While this approach is powerful and shows promise for effective adaptation to our use case, other works have emerged with simpler architectures. A notable example is Medical MAISI \cite{guoMAISIMedicalAI2024}. Instead of operating on patches, MAISI compresses full volumes directly. It then uses a latent diffusion model based on a time-conditional U-Net, conditioned on body region and voxel spacing, following the general latent-space design introduced by Rombach et al.~\cite{rombachHighResolutionImageSynthesis2022} and Pinaya et al.~\cite{pinayaBrainImagingGeneration2022}. Their autoencoder is trained on over 37,000 CT and 17,000 MRI scans, enabling strong generalization to unseen data. Without any additional fine-tuning, it matches, and in some cases surpasses the performance of dedicated autoencoders that were pretrained for hundreds of GPU hours. They additionally train their latent diffusion model on 10,277 CT volumes and, by using tensor-splitting parallelism, achieve image generation at resolutions of up to $512\times512\times768$. Similarly to 3D MedDiffusion, MAISI supports spatial conditioning via a ControlNet. Recently, MAISI-v2 \cite{zhaoMAISIv2Accelerated3D2025} was introduced, following the same principle as the original work with several technical improvements, including the use of rectified flow to accelerate sampling. In this work, we use MAISI for our post-treatment brain MRI prediction task due to its simple yet effective architecture, its ability to accommodate our multi-modal conditioning needs, and the availability of publicly shared pretrained autoencoder weights.

\section{Methods}
\label{sec:methods}

\subsection{Data}

We used the public SAILOR dataset \cite{hovdenBrainTumorMRI2023}, consisting of 27 patients with high-grade glioblastoma treated at Oslo University Hospital. All patients underwent debulking surgery followed by the Stupp protocol \cite{stuppRadiotherapyConcomitantAdjuvant2005}: 60 Gy of fractionated radiotherapy with concomitant temozolomide, followed by six cycles of adjuvant temozolomide chemotherapy. Each patient has a pretreatment planning MRI, a radiotherapy (RT) dose map, and between 3 and 19 follow-up scans acquired at regular intervals during and after treatment. After removing two patients with corrupted RT dose maps, we retained 257 MRI scans from 25 patients.

For each patient, the dataset includes several structural MRI sequences: T1-weighted (T1w), T1-weighted contrast-enhanced (T1Gd), and T2 Fluid Attenuated Inversion Recovery (T2-FLAIR), among others. In this work, we use only T1-weighted images to simplify the initial 3D implementation. The dataset comes preprocessed with N4 bias-field correction, Rician denoising, FSL skull stripping, 1mm voxel isotropic resampling, intra-patient rigid registration, longitudinal intensity normalization via Piecewise Linear Histogram Matching (PLHM), and affine registration to the MNI152 template \cite{hovdenBrainTumorMRI2023}. Intensities were additionally normalized per volume such that the $0.1$st to $99.9$th percentile values were scaled to $[0,1]$. For training and evaluation, the raw image dimensions of $193 \times 229 \times 193$ were center-cropped to $160 \times 224 \times 192$. This resolution was chosen to satisfy compatibility with the downsampling factors of the autoencoder and U-Net, while closely fitting the brain volume to avoid unnecessary computation on background voxels. The dataset was split into 21 training patients, 2 validation patients, and 2 test patients, following the same split as the 2D baseline \cite{huismanRectifiedFlowbasedPrediction2026}.

\subsection{Development stages}

\begin{figure*}[!t]
\centering
\includegraphics[width=\textwidth]{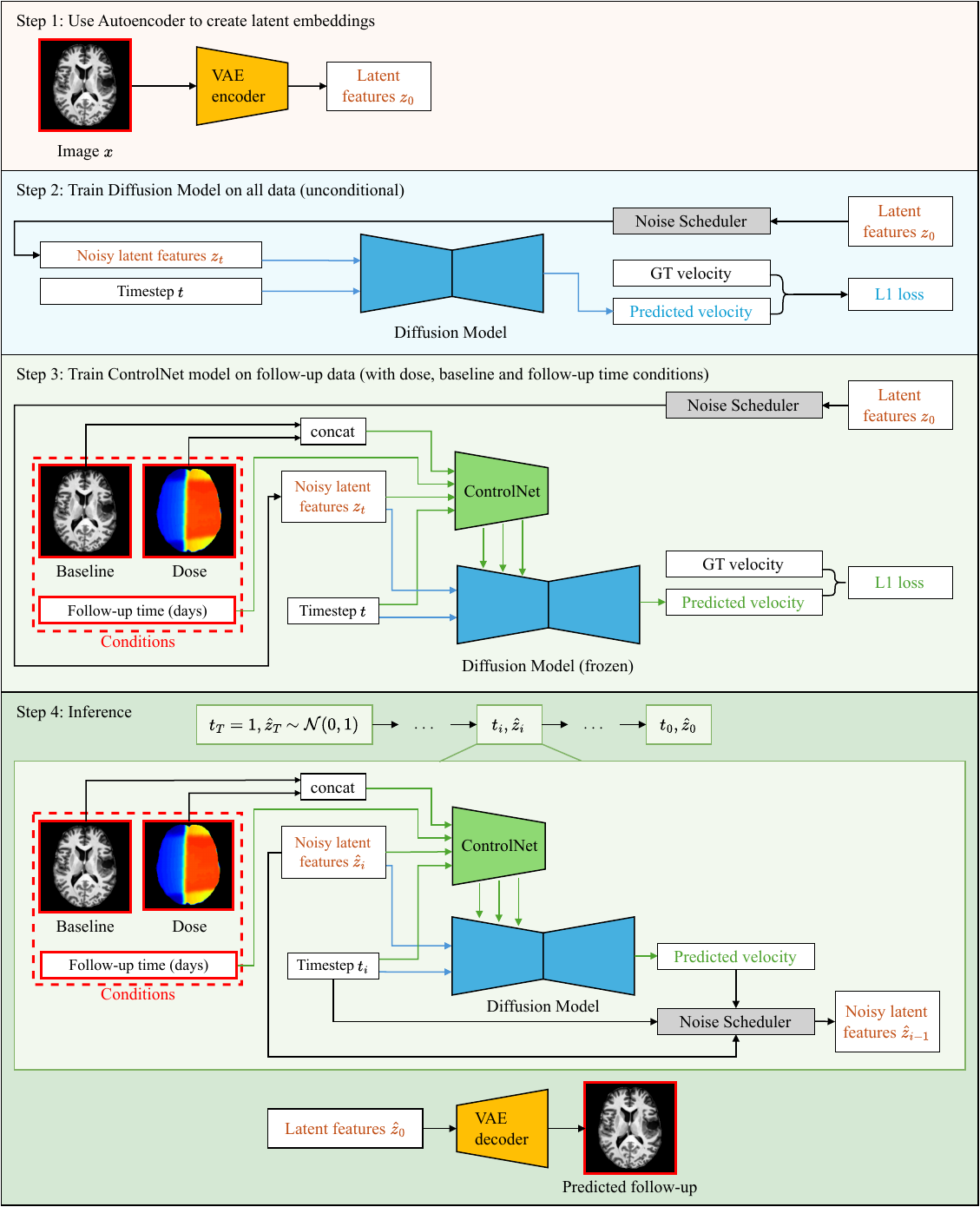}
\caption{Pipeline and model architecture. The figure shows a single training iteration of the latent diffusion model and the ControlNet. During training, the noise scheduler adds noise to the clean latent via linear interpolation, following Eq.~\ref{eq:linear_interpolation}. GT velocity denotes ground truth velocity. During inference, the noise scheduler keeps track of the previous timestep and updates the noisy latent using the predicted velocity, following Eq.~\ref{eq:inference_latent_update}.}
\label{fig:architecture}
\end{figure*}

We follow the same development stages as the MAISI framework \cite{guoMAISIMedicalAI2024}, with full details on our adaptation described below. The overall pipeline is depicted in Fig.~\ref{fig:architecture}.

\subsubsection{Latent Compression}
\label{sec:latent_compression}
Let $x$ denote a single T1-weighted brain scan of our dataset, with $x \in \mathbb{R}^{1 \times H \times W \times D}$, where the first dimension is the image channel and $H$, $W$, and $D$ are the spatial dimensions in image space. The MAISI autoencoder, with encoder $\mathcal{E}nc$ and decoder $\mathcal{D}ec$, maps each scan to a compact latent feature representation $z_0 = \mathcal{E}nc(x) \in \mathbb{R}^{C \times h \times w \times d}$, where $C$ is the number of latent channels and $h$, $w$, and $d$ are the spatial dimensions in latent space. The notation $z_0$ is used to reflect the target result of the iterative denoising process done by the latent diffusion model in the next development stage.

As a foundation model, the MAISI autoencoder \cite{guoMAISIMedicalAI2024} is sufficiently robust to be utilized directly, without fine-tuning. Before diffusion training, the model's encoder is applied once to all scans in the training set. We denote the resulting set of latent features by $\mathcal{Z}$. All later generative modeling is performed in this latent space.

\subsubsection{Latent Diffusion with Rectified Flow}
A diffusion U-Net $v_\theta$, following the design of Rombach et al.~\cite{rombachHighResolutionImageSynthesis2022} and Pinaya et al.~\cite{pinayaBrainImagingGeneration2022}, is used as the generative backbone. The model is trained on latent features $z_0 \in \mathcal{Z}$. Training on all baseline and follow-up latents yields an unconditional prior that captures realistic T1 brain structure and appearance, providing a stable generative foundation for subsequent treatment-based conditioning.

In standard diffusion (DDPM) \cite{hoDenoisingDiffusionProbabilistic2020a} as well as latent diffusion, a forward Markov process is defined by iteratively adding noise to a clean sample $z_0$, resulting in a sequence of noisy samples $z_1, \dots, z_T$, where $z_T$ approaches pure noise. A denoising network is then trained to parameterize the corresponding reverse Markov process, commonly via noise prediction at a particular time step $t$. In this work, we instead adopt rectified flow \cite{liuFlowStraightFast2022}, which learns a velocity field that defines an ordinary differential equation (ODE) transporting samples from noise to data. When sampling is viewed as numerically transporting noise to data, DDPM trajectories are generally not straight, so accurate generation often requires many small steps to control discretization error. Rectified flow explicitly biases this ODE toward near-straight paths, enabling accurate integration with relatively few steps. Previous works have shown that rectified flow can achieve inference speeds 33x \cite{zhaoMAISIv2Accelerated3D2025}, sometimes even 250x \cite{zhaoMAISIv2Accelerated3D2025} faster than DDPM-based sampling. Our training setup follows MAISI \cite{guoMAISIMedicalAI2024}, replacing DDPM with rectified flow as in MAISI-v2 \cite{zhaoMAISIv2Accelerated3D2025}, and omitting the body-region and voxel-spacing conditions. During training, we sample Gaussian noise $\epsilon \sim \mathcal{N}(0,I)$ and a normalized timestep $t \sim \mathrm{Uniform}(\mathcal{T})$, where $\mathcal{T}$ is $\{\frac{1}{T}, \frac{2}{T}, \dots, 1\}$ remapped via the resolution-aware timestep shifting introduced in Stable Diffusion 3 \cite{esserScalingRectifiedFlow2024}. This method shifts timesteps toward $1$ for larger volumes, allocating more timesteps to the early denoising stages. The noisy latent is then constructed by linear interpolation:
\begin{equation}
\label{eq:linear_interpolation}
    z_t = (1-t) z_0 + t\,\epsilon.
\end{equation}
The network $v_\theta$ is trained to predict the velocity target $v^\star = z_0 - \epsilon$. Specifically, we minimize the L1 objective
\begin{equation*}
    \mathcal{L}_{\mathrm{RF}}(\theta)
    = \mathbb{E}_{z_0, \epsilon, t}\left[ \left\| v_\theta(z_t, t) - (z_0 - \epsilon) \right\|_1 \right].
\end{equation*}

\subsubsection{ControlNet Conditioning}
After training our denoising backbone, we use a ControlNet \cite{controlnet} $\mathcal{C}_\phi$ to condition the generation on treatment information. The ControlNet copies the encoder and middle blocks of the pretrained U-Net into a trainable parallel branch and injects features back into the encoder through zero-initialized convolutions at multiple levels. This preserves the pretrained model's latent-space generative capabilities, while learning task-specific control signals.

Each supervised sample is composed of a target follow-up scan $x_{\mathrm{fu}}$, the corresponding baseline scan $x_{\mathrm{bl}}$, the radiotherapy dose map $x_{\mathrm{dose}}$, and the time since the start of treatment (in days) $d$, where $x_{\mathrm{fu}}, x_{\mathrm{bl}}, x_{\mathrm{dose}} \in \mathbb{R}^{1 \times H \times W \times D}$. The conditioning network includes a 3D convolutional encoder $f$ to handle spatial inputs. The baseline scan and dose map are concatenated along the channel dimension to form $\mathrm{concat}(x_{\mathrm{bl}}, x_{\mathrm{dose}}) \in \mathbb{R}^{2 \times H \times W \times D}$, and $f$ maps this input to a spatial condition $c_s = f(\mathrm{concat}(x_{\mathrm{bl}}, x_{\mathrm{dose}})) \in \mathbb{R}^{C \times h \times w \times d}$, matching latent feature dimensions. The follow-up time $d$ is encoded with a timestep embedding MLP. Given a noisy target latent $z_t$ and diffusion timestep $t$, the ControlNet produces a conditioning signal $c = \mathcal{C}_\phi(z_t, t, c_s, d)$, which is injected into the frozen U-Net at each individual encoder block stage. The conditioned velocity is then predicted as $\hat{v} = v_\theta(z_t, t, c)$.

During inference, generation starts from Gaussian noise $z_T \sim \mathcal{N}(0, I)$ and timesteps follow the shifted schedule $\mathcal{T}$ in reverse order, from $t_T = 1$ down to $t_1$. At each step $i$ in $T, \dots, 1$, the ControlNet produces a conditioning signal $c = \mathcal{C}_\phi(z_i, t_i, c_s, d)$, which is then used to predict the conditioned velocity $\hat{v} = v_\theta(z_i, t_i, c)$. The latent is then updated as
\begin{equation}
    \label{eq:inference_latent_update}
    z_{i-1} = z_i + (t_i - t_{i-1})\,\hat{v},
\end{equation}
with $t_0 = 0$. The resulting latent at the final step is denoted by $z_0$ and decoded into image space as $\hat{x}_{\mathrm{fu}} = \mathcal{D}ec(z_0)$. The training procedure and inference are shown in detail in Algorithms~\ref{alg:training} and~\ref{alg:inference}.

\begin{algorithm}[!t]
\caption{ControlNet Training}
\label{alg:training}
\small
\KwIn{Training set $\mathcal{D}$ of tuples $(z_0, x_{\mathrm{bl}}, x_{\mathrm{dose}}, d)$, where $z_0$ is the precomputed target follow-up latent; frozen U-Net $v_\theta$; ControlNet $\mathcal{C}_\phi$, including spatial encoder $f$; timestep schedule $\mathcal{T} = \{t_1, \dots, t_T\}$}
\For{each $(z_0, x_{\mathrm{bl}}, x_{\mathrm{dose}}, d)$ in $\mathcal{D}$}{
    $t \sim \mathrm{Uniform}(\mathcal{T})$\;
    $\epsilon \sim \mathcal{N}(0, I)$\;
    $c_s \gets f(\mathrm{concat}(x_{\mathrm{bl}}, x_{\mathrm{dose}}))$\;
    $z_t \gets (1-t)\, z_0 + t\,\epsilon$\;
    $c \gets \mathcal{C}_\phi(z_t, t, c_s, d)$\;
    $\hat{v} \gets v_\theta(z_t, t, c)$\;
    $\mathcal{L} \gets \left\|\hat{v} - (z_0 - \epsilon)\right\|_1$\;
    Update $\phi$ via $\nabla_\phi \mathcal{L}$\;
}
\end{algorithm}

\begin{algorithm}[!t]
\caption{ControlNet Inference}
\label{alg:inference}
\small
\KwIn{Baseline scan $x_{\mathrm{bl}}$, dose map $x_{\mathrm{dose}}$, target time $d$, timestep schedule $\mathcal{T} = \{t_1, \dots, t_T\}$}
\KwOut{Predicted follow-up scan $\hat{x}_{\mathrm{fu}}$}
$c_s \gets f(\mathrm{concat}(x_{\mathrm{bl}}, x_{\mathrm{dose}}))$\;
$z_T \sim \mathcal{N}(0, I)$\;
\For{$i = T$ \KwTo $1$}{
    $c \gets \mathcal{C}_\phi(z_i, t_i, c_s, d)$\;
    $\hat{v} \gets v_\theta(z_i, t_i, c)$\;
    $z_{i-1} \gets z_i + (t_i - t_{i-1})\,\hat{v}$\;
}
$\hat{x}_{\mathrm{fu}} \gets \mathcal{D}ec(z_0)$\;
\Return{$\hat{x}_{\mathrm{fu}}$}
\end{algorithm}

\subsection{Evaluation}
\label{sec:evaluation}

We evaluated predictions using several categories of metrics designed to comprehensively capture different aspects of prediction quality.

First, we assess general image quality using Mean Squared Error (MSE), Peak Signal-to-Noise Ratio (PSNR), and Structural Similarity Index Measure (SSIM) between predicted and real follow-up scans. While MSE and PSNR are sensitive to intensity shifts, SSIM evaluates luminance, contrast, and structural similarity, making it more robust to global intensity variations and more aligned with perceptual quality. We report 3D metrics (3D SSIM, 3D PSNR, 3D MSE) computed on full volumes at our native training resolution. However, to enable a fair comparison with the 2D baseline \cite{huismanRectifiedFlowbasedPrediction2026}, we also report 2D variants of all metrics, calculated on slices between 42 and 140, matching the baseline. To ensure consistency, the images are cropped to the baseline’s image resolution, as the baseline directly operates in image space without autoencoder compression. Due to the matched resolutions, the metrics cannot be inflated by background voxels, allowing for direct comparison. We note that SSIM in particular is different in 3D and 2D. The 3D variant applies a volumetric sliding window kernel that computes local statistics over three-dimensional neighborhoods and thus captures through-plane structural correlations, whereas the 2D variant applies a standard 2D kernel per slice independently.

To assess anatomical consistency, we compute Dice scores on tissue segmentations. Both predicted and real follow-up scans were segmented into cerebrospinal fluid (CSF), gray matter, and white matter using FSL FAST \cite{zhangSegmentationBrainMR2001, jenkinsonFSL2012a}. Unlike previous 2D work which used two classes (tissue and CSF), we evaluate three classes to provide a finer-grained assessment of tissue differentiation. To visually compare the segmentation results, we plot real and predicted follow-up segmentation maps, including an error map highlighting regions of discrepancy. In addition, recognizing the cognitive impacts of radiation, we evaluate the absolute and relative prediction errors on whole intact hippocampus volumes extracted using SynthSeg \cite{billotSynthSegSegmentationBrain2023}.

We also perform deformation-based morphometry (DBM) to assess how accurately the model captures tissue volume changes over time. First, non-linear image registration is used to estimate a displacement field relating the baseline and follow-up scans. From this field, voxel-wise Jacobian determinants are computed to quantify local volume change. In the 2D baseline study, registration was performed using FSL FNIRT \cite{jenkinsonFSL2012a}, a classical optimization-based method applied to image pairs. In contrast, we use SynthMorph \cite{hoffmannSynthMorphLearningContrastinvariant2022}, a pretrained deep learning-based registration framework that is contrast-invariant, anatomically informed, and typically faster for volumetric data. Jacobian determinant maps are then computed from the displacement fields using Advanced Normalization Tools (ANTs) \cite{tustisonANTsXEcosystemQuantitative2021}, applying a natural log transformation to the results. In these maps, negative values indicate local tissue contraction, whereas positive values indicate local expansion. Because the natural logarithm is used, a log Jacobian value of $x$ corresponds to a local volume change of exactly $(e^x - 1) \times 100\%$.

Finally, we perform a longitudinal analysis in which samples were grouped into follow-up time bins (in days after the start of treatment). Dice scores were then compared across bins to assess whether prediction accuracy changed with follow-up duration. 

\subsection{Implementation Details}

We trained the latent diffusion model (U-Net) for 1000 epochs with a batch size of $1$ on an NVIDIA RTX 4080 Super GPU. We use an Adam optimizer with automatic mixed precision, gradient scaling, and a quadratically decaying learning rate starting at $1 \times 10^{-4}$. The model was trained for approximately 48 hours. The ControlNet is trained for 2000 epochs with a batch size of $16$ on an NVIDIA H100 GPU for approximately 17 hours. We use an AdamW optimizer with automatic mixed precision, gradient scaling, and a linearly decaying learning rate starting at $1 \times 10^{-4}$. We save model checkpoints and measure 3D SSIM on the validation set every 10 epochs. The final model is selected based on the best validation score.

The shifting transformation of the timestep schedule is parameterized by the size ratio
\begin{equation*}
    r = \left(\frac{m}{n}\right)^{1/3},
\end{equation*}
where $m = h \cdot w \cdot d$ denotes the latent volume size and $n$ is a reference volume size. Using the reference size adopted in the MONAI implementation of rectified flow, $n = 32^{3}$ \cite{cardosoMONAIOpensourceFramework2022}, and our latent resolution of $40 \times 56 \times 48$, we obtain $r \approx 1.49$. The same resolution-aware timestep shifting is applied in both training and inference, with $T=1000$ during training and $T=10$ during inference. This shifting is motivated by the observation that higher-resolution volumes require more noise to destroy their signal: by matching per-voxel signal uncertainty across resolutions, a timestep $t$ at the base resolution can be mapped to a shifted timestep $t'$ at a higher resolution while preserving the same degree of uncertainty \cite{esserScalingRectifiedFlow2024}.

\section{Results}
\label{sec:results}

\subsection{Quantitative Performance}

Table~\ref{tab:metrics} summarizes image quality metrics for our 3D model and the 2D baseline \cite{huismanRectifiedFlowbasedPrediction2026}. Our model achieves a 3D SSIM of $0.941 \pm 0.061$ and a 2D SSIM of $0.905 \pm 0.092$, compared to $0.88 \pm 0.08$ for the 2D baseline. 2D PSNR also improves from $22.82 \pm 4.02$ to $25.82 \pm 5.49$ dB, and 2D MSE decreases from $0.008 \pm 0.007$ to $0.0062 \pm 0.0085$.

\begin{table}[!t]
\centering
\begin{threeparttable}
\caption{Image quality comparison between our 3D model and the 2D baseline.}
\label{tab:metrics}
\begin{tabular}{clcc}
\toprule
\textbf{Setting} & \textbf{Metric} & \textbf{Ours} & \textbf{2D Baseline} \\
\midrule
\multirow{3}{*}{3D$^a$} 
 & SSIM $\uparrow$ & 0.941 $\pm$ 0.061 & -- \\
 & PSNR $\uparrow$ & 27.67 $\pm$ 5.29 & -- \\
 & MSE $\downarrow$ & 0.0039 $\pm$ 0.0053 & -- \\
\midrule
\multirow{3}{*}{2D$^b$} 
 & SSIM $\uparrow$ & \textbf{0.905} $\pm$ 0.092 & 0.88 $\pm$ 0.08 \\
 & PSNR $\uparrow$ & \textbf{25.82} $\pm$ 5.49 & 22.82 $\pm$ 4.02 \\
 & MSE $\downarrow$ & \textbf{0.0062} $\pm$ 0.0085 & 0.008 $\pm$ 0.007 \\
\bottomrule
\end{tabular}
\begin{tablenotes}
    \footnotesize
    \item[a] Native training resolution (see Section~\ref{sec:evaluation}).
    \item[b] 2D baseline resolution, slice-wise (see Section~\ref{sec:evaluation}).
\end{tablenotes}
\end{threeparttable}
\end{table}

Mean Dice score across CSF, gray matter, and white matter is $0.811 \pm 0.119$, with per-class scores of $0.761 \pm 0.140$ for CSF, $0.772 \pm 0.132$ for gray matter, and $0.901 \pm 0.085$ for white matter. Log Jacobian determinant analysis yields a mean absolute error of $0.037 \pm 0.025$, corresponding to a mean predicted local volume change of $(e^{0.037} - 1) \times 100\% \approx 3.82\%$ on average.

\subsection{Qualitative Results}

\begin{figure*}[!t]
\centering
\includegraphics[width=\textwidth]{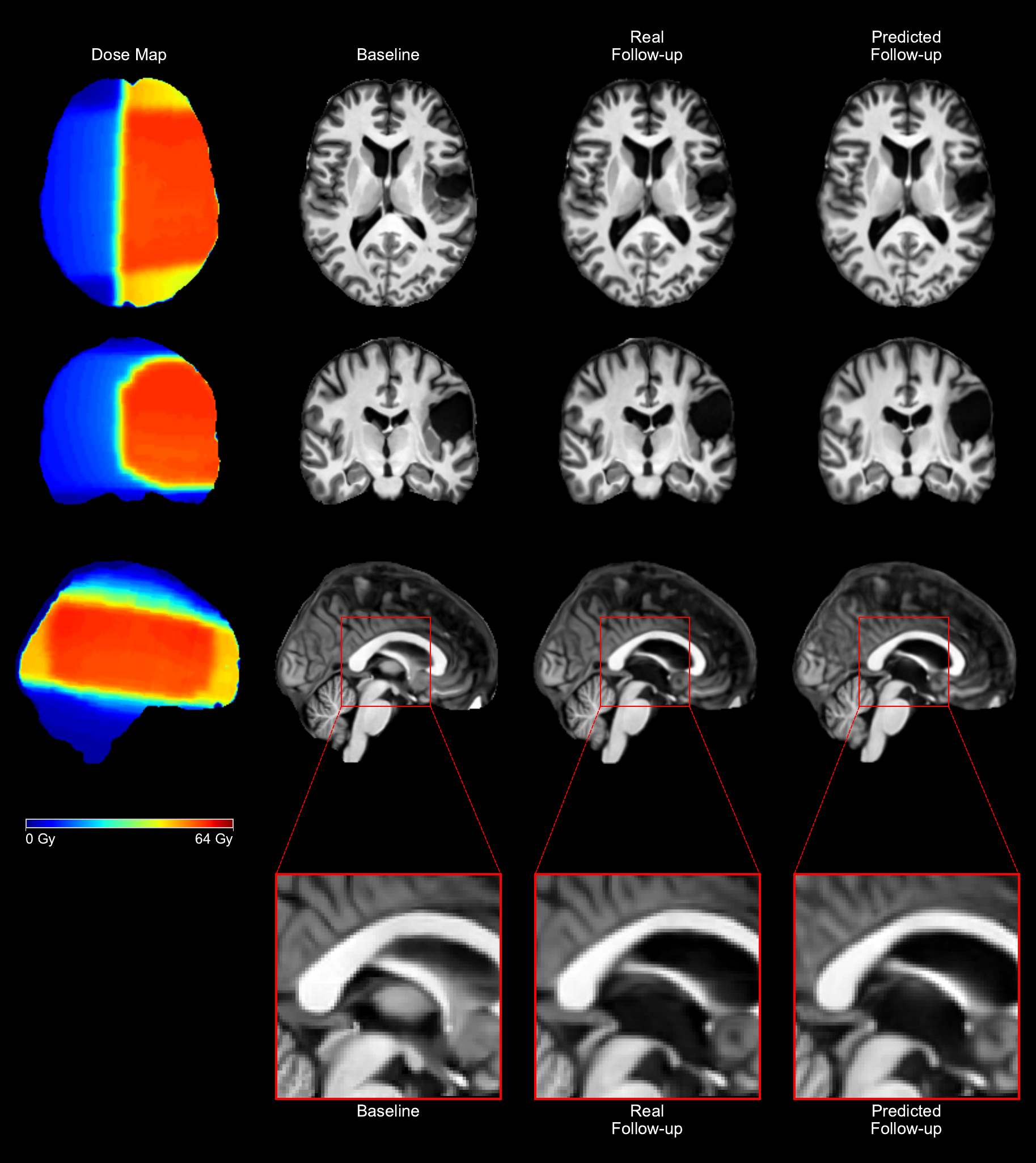}
\caption{Prediction for a single test patient. Columns show the radiation dose map (0--64\,Gy), baseline T1 scan, real follow-up, and predicted follow-up. Rows show the axial, coronal, and sagittal planes, with a zoomed-in crop of the sagittal view highlighting structural detail in the corpus callosum region.}
\label{fig:prediction}
\end{figure*}

Fig.~\ref{fig:prediction} shows a 3D prediction for a single test patient, displayed across axial, coronal, and sagittal planes alongside the corresponding dose map, baseline, and real follow-up. The predicted follow-up closely resembles the ground truth in overall brain morphology and intensity distribution. Patient-specific anatomy from the baseline is preserved, and no noise or hallucinated structures are introduced. A zoomed-in crop of the sagittal view confirms structural continuity at fine anatomical detail, but reveals regions with slight blurriness in the predicted follow-up compared to the ground truth, particularly in superior portions of the zoomed region.

\normalcolor

\subsection{Segmentation Analysis}

\begin{figure}[!t]
\centering
\includegraphics[width=\columnwidth]{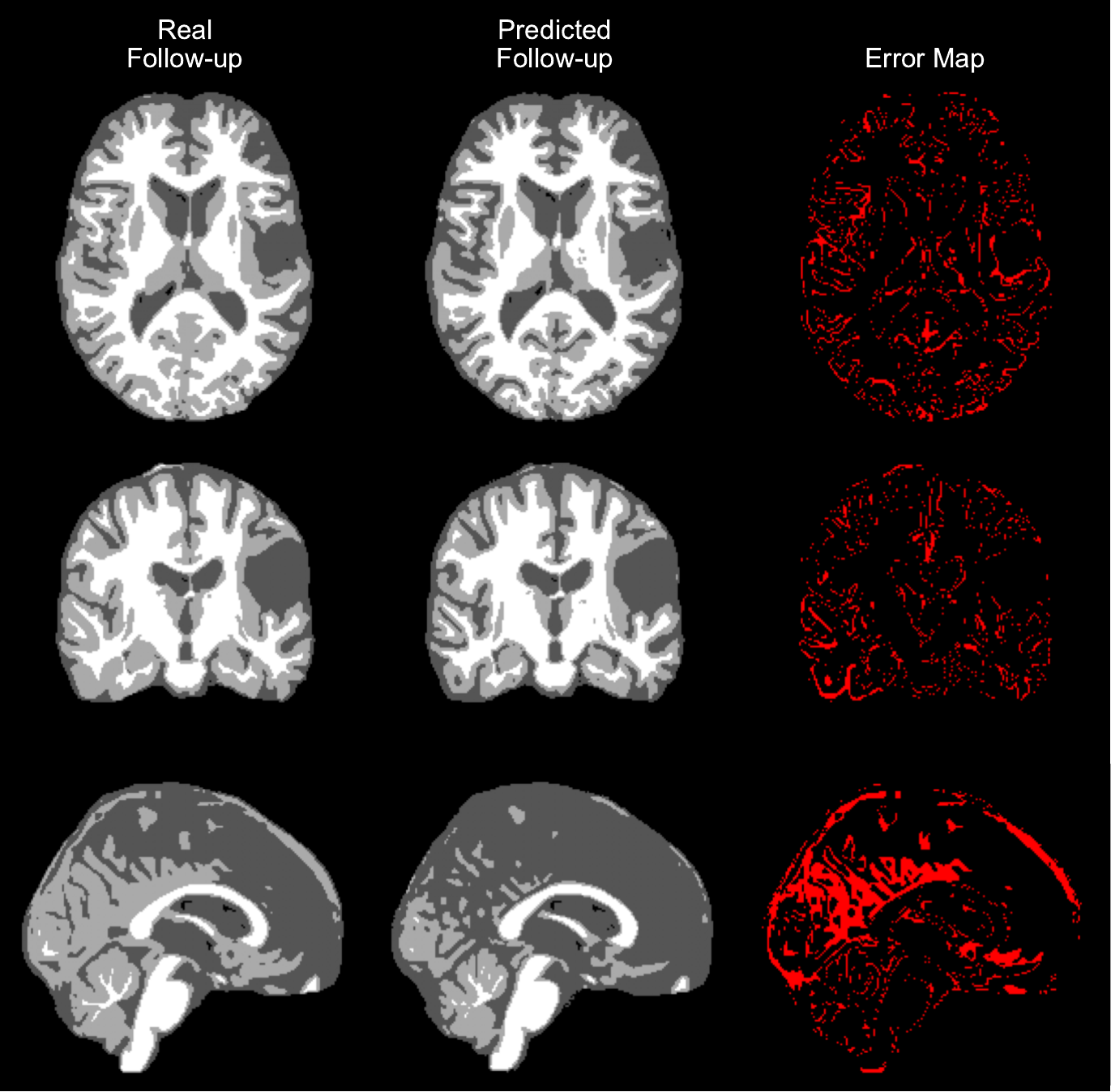}
\caption{Tissue segmentation comparison for a test patient. Columns show real follow-up segmentation, predicted follow-up segmentation, and an error map (red indicates label disagreement). Rows show axial, coronal, and sagittal planes. Gray levels correspond to CSF (dark), gray matter (mid), and white matter (light).}
\label{fig:segmentation}
\end{figure}

Fig.~\ref{fig:segmentation} shows tissue segmentation maps for a test sample, comparing real and predicted follow-up segmentation maps across axial, coronal, and sagittal planes. An error map in the third column highlights voxels where the predicted segmentation label differs from ground truth. Mean Dice score across CSF, gray matter, and white matter is $0.811 \pm 0.119$, with white matter achieving the highest per-class score ($0.901 \pm 0.085$), followed by gray matter ($0.772 \pm 0.132$) and CSF ($0.761 \pm 0.140$).

Fig.~\ref{fig:segmentation} shows that segmentation performance degrades in regions where pathology such as edema is visually significant, for instance around the resection cavity on the right side of the brain. Additionally, the bottom error map on the figure shows widespread misclassifications of gray matter as CSF and vice versa.

\subsection{Deformation Analysis}

\begin{figure}[!t]
\centering
\includegraphics[width=\columnwidth]{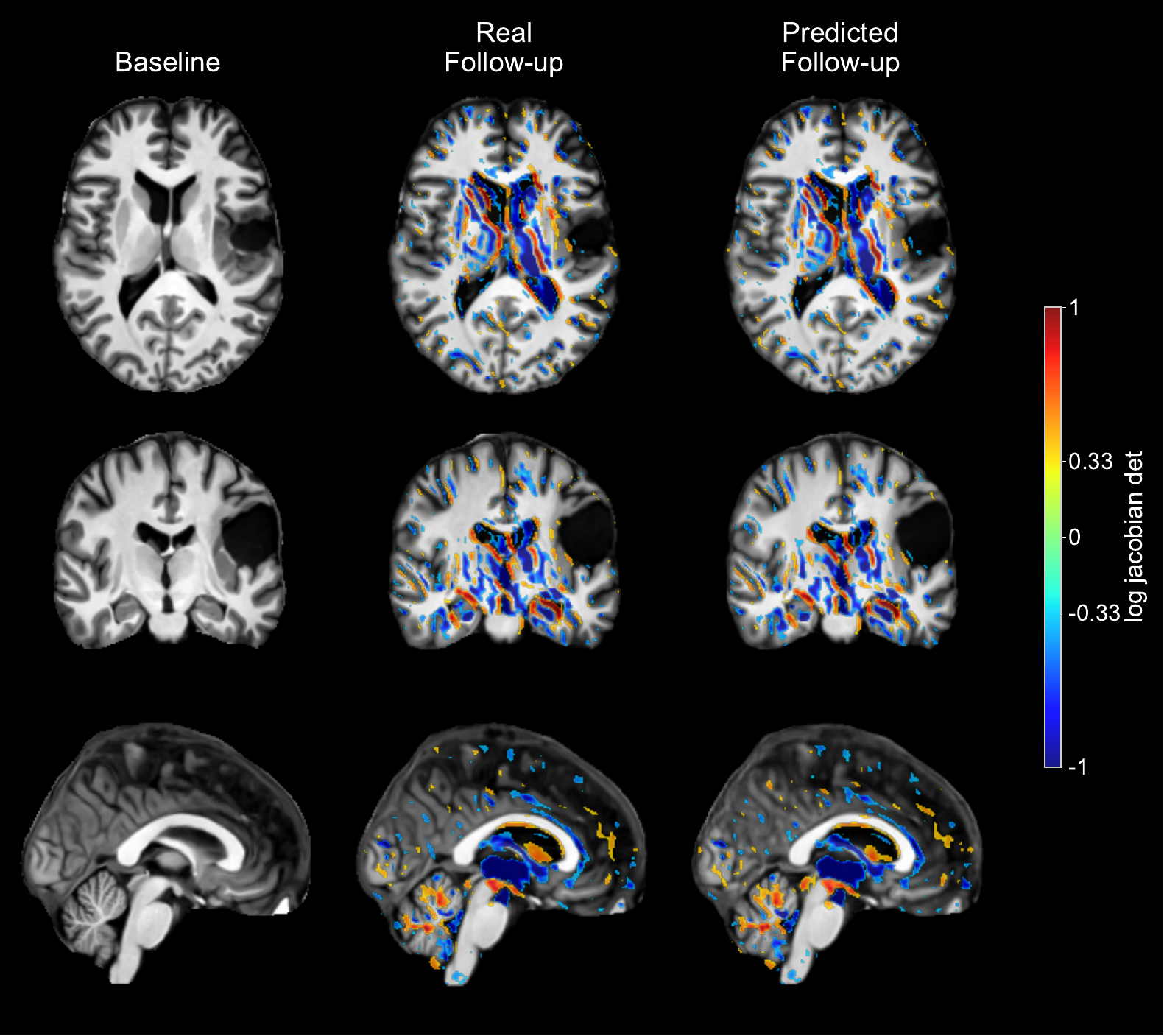}
\caption{Log Jacobian determinant maps for a test patient. Columns show the baseline T1 scan, the log Jacobian map derived from registering baseline to real follow-up, and the log Jacobian map from baseline to predicted follow-up. Rows show axial, coronal, and sagittal planes. Warm colors (positive values) indicate local expansion; cool colors (negative values) indicate local contraction.}
\label{fig:jacobian}
\end{figure}

Fig.~\ref{fig:jacobian} shows log Jacobian determinant maps for a test patient, comparing real and predicted deformation fields. The maps show strong spatial agreement between real and predicted volume changes. Regions of ventricular expansion (positive log Jacobian, warm colors) and cortical contraction (negative log Jacobian, cool colors) are consistently reproduced in the predicted maps. The mean absolute error of $0.037 \pm 0.025$ across the evaluated dataset confirms this qualitative agreement quantitatively.

\subsection{Longitudinal Performance}

\begin{figure}[!t]
\centering
\includegraphics[width=\columnwidth]{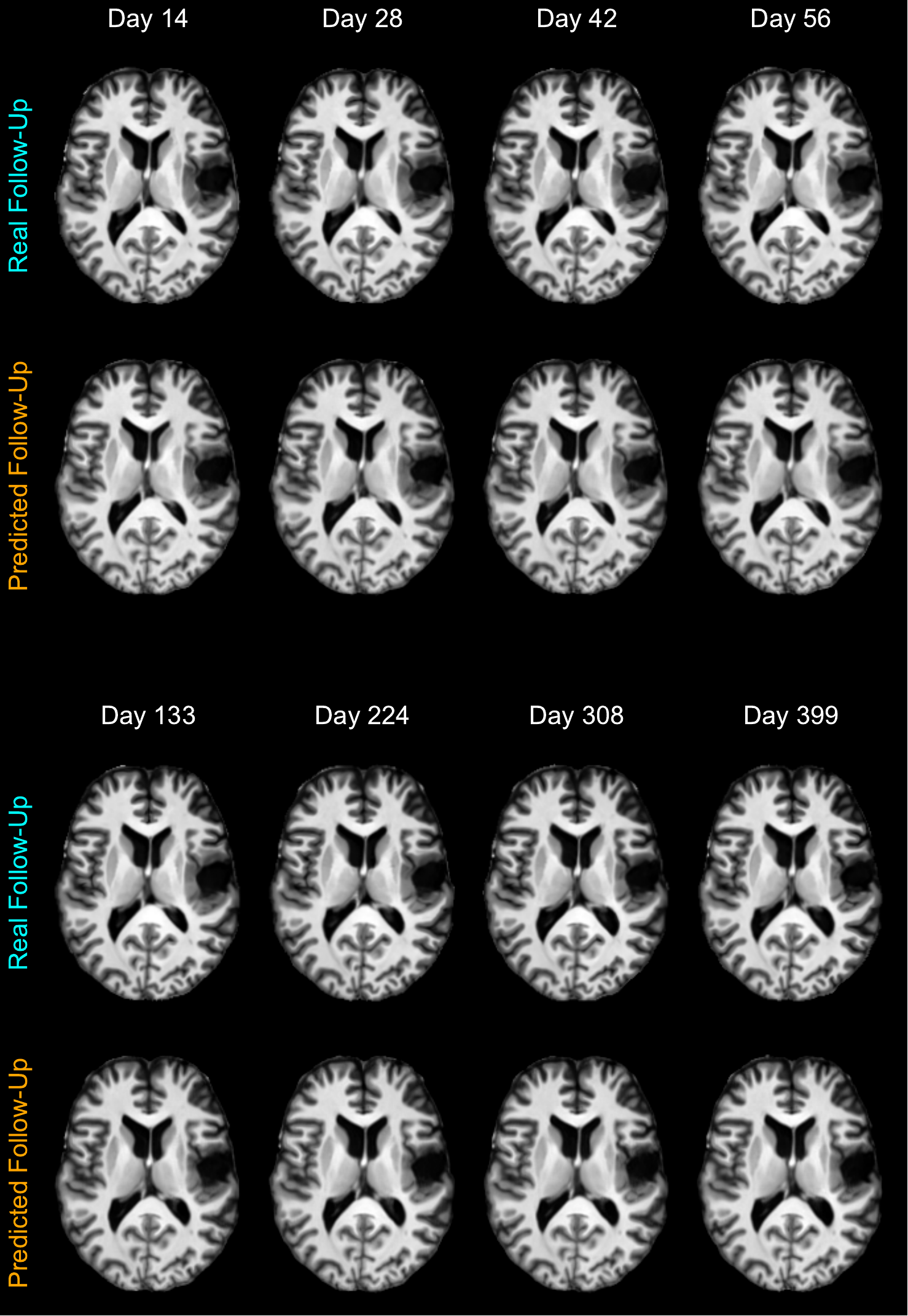}
\caption{Real and predicted follow-up scans for a single test patient at Days 14, 28, 42, 56, 133, 224, 308, and 399 after the start of treatment. Each group shows two rows: real follow-up (cyan label) and predicted follow-up (orange label). Each column corresponds to one timepoint.}
\label{fig:longitudinal}
\end{figure}

Fig.~\ref{fig:longitudinal} shows real and predicted follow-up scans for a single test patient at several timepoints after the start of treatment. The model produces predictions that closely track the real follow-up scans across all timepoints. Progressive ventricular enlargement, visible in the real scans at later timepoints, is reproduced in the corresponding predictions. No visible degradation in prediction quality is observed at later timepoints.

\begin{figure}[!t]
\centering
\includegraphics[width=\columnwidth]{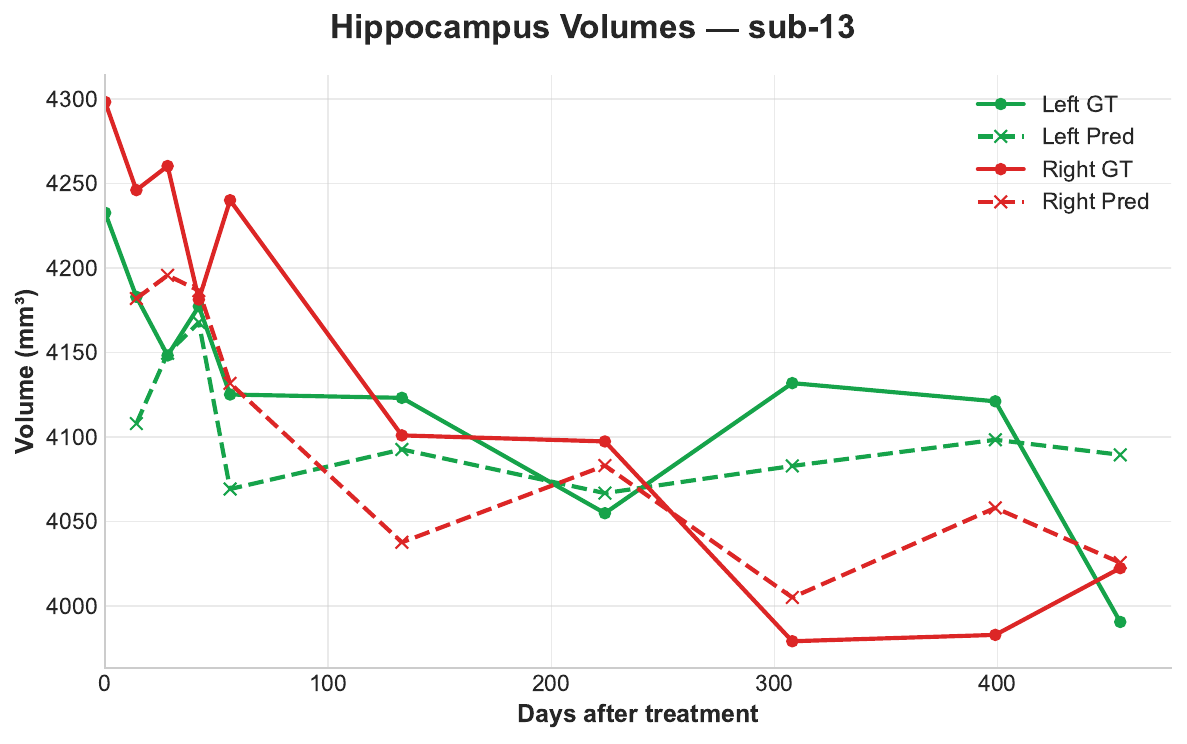}
\caption{Predicted and real hippocampus volumes over time for test patient sub-13. Solid lines show ground truth; dashed lines show model predictions. Green and red indicate left and right hippocampus, respectively.}
\label{fig:hippocampus}
\end{figure}

Table~\ref{tab:hippocampus} reports the mean absolute and relative hippocampus volume prediction errors for both test patients. Fig.~\ref{fig:hippocampus} shows predicted and real hippocampus volumes over time. The predicted trajectories follow the general temporal trend of the real follow-ups for both left and right hippocampus, with relative differences remaining below 3.5\% across all timepoints.

\begin{table}[!t]
\centering
\caption{Hippocampus Volume Prediction Error (Mean Across All Follow-Up Timepoints)}
\label{tab:hippocampus}
\resizebox{\columnwidth}{!}{%
\begin{tabular}{lcccc}
\toprule
& \multicolumn{2}{c}{\textbf{Left Hippocampus}} & \multicolumn{2}{c}{\textbf{Right Hippocampus}} \\
\cmidrule(lr){2-3} \cmidrule(lr){4-5}
\textbf{Patient} & \textbf{Abs.\ Error (mm$^3$)} & \textbf{Rel.\ Error (\%)} & \textbf{Abs.\ Error (mm$^3$)} & \textbf{Rel.\ Error (\%)} \\
\midrule
sub-01 & 100.96 & 1.68 & 208.68 & 3.40 \\
sub-13 & 39.43  & 0.96 & 47.14  & 1.14 \\
\midrule
\textit{Mean} & \textit{70.20} & \textit{1.32} & \textit{127.91} & \textit{2.27} \\
\bottomrule
\end{tabular}%
}
\end{table}

\section{Discussion}
\label{sec:discussion}

This work demonstrates that full-volume post-radiotherapy brain MRI prediction is possible using a treatment-aware 3D latent diffusion framework. Unlike slice-wise approaches, the proposed model operates directly on volumetric anatomy and incorporates the spatial radiation dose distribution as a conditioning signal. This is important because radiation effects are inherently three-dimensional: local tissue response depends not only on the dose received at a single location, but also on the surrounding anatomy, regional connectivity, and distributed dose gradients across the brain. The improved image-similarity scores relative to the prior 2D framework, together with strong tissue segmentation agreement, hippocampal volume accuracy, and deformation-map consistency, indicate that the model learns clinically meaningful anatomical structure rather than merely producing visually plausible images.

In image quality scores, the 3D model improves upon the 2D baseline across all reported metrics, but the improvement is particularly notable for SSIM (2D: 0.91 vs. 0.88 for the baseline, 3D: 0.94), which evaluates structural similarity and is less sensitive to global intensity shifts than MSE or PSNR. This suggests the model's ability to maintain spatial coherence across the volume, an expected behavior from a dedicated 3D model. Furthermore, the log Jacobian determinant analysis provides evidence that the model predicts biologically plausible tissue changes rather than merely matching appearance. With a mean absolute error of $0.037$ in log Jacobian, the predicted volume changes closely approximate those of real patients, with an error of only about 3.82\%. This matters for clinical applications where considering the spatial distribution of tissue change is more relevant than pixel-level similarity alone. Similarly, hippocampus volume predictions remain within 1.0--3.4\% of ground truth values, indicating that the model preserves region-specific volumetric accuracy despite operating on a highly compressed latent representation.

Qualitative assessment did not always align with quantitative image-similarity metrics. In some cases, predictions with relatively low metric values nevertheless preserved plausible contrast, anatomy, and overall structural organization, whereas some predictions with high metric values appeared less faithful on visual inspection. This mismatch likely reflects the sensitivity of MSE, PSNR, and SSIM to small voxel-wise intensity differences, residual misregistration, boundary shifts, and smoothing effects, which do not necessarily correspond to clinically meaningful anatomical plausibility. It may also reflect the intrinsic uncertainty of longitudinal prediction, where multiple future anatomical trajectories can be biologically plausible although only one observed follow-up is available for metric-based comparison. Therefore, we complement conventional image-similarity metrics with tissue segmentation, hippocampal volumetry, and deformation-based analyses, which provide a more anatomically and biologically informed assessment of prediction quality.

\subsection{Clinical Implications}

The central implication of this work is that patient-specific post-radiotherapy brain MRI can be predicted using pretreatment imaging, the planned radiation dose distribution, and a target follow-up time. This represents a shift from describing radiation effects retrospectively to modeling them prospectively at the level of the individual patient, which opens perspectives for personalized radiotherapy approaches. Rather than producing only scalar risk estimates or population-level dose-response summaries, the proposed framework generates a full 3D virtual follow-up scan that preserves anatomical context and can be evaluated with the same image-analysis tools used for real longitudinal MRI.

This capability is directly relevant for radiotherapy planning. Current treatment planning relies primarily on target coverage, dose constraints, and population-level normal tissue assumptions implying that all brains react to equal doses in equal ways. A treatment-aware generative model adds a complementary layer: it predicts the spatial anatomical consequences of a planned dose distribution in the individual patient's brain. Clinicians and researchers could therefore examine predicted changes in vulnerable structures such as the hippocampus, cortex, deep gray matter, and white matter, rather than evaluating dose maps alone. Because the output is a complete 3D MRI volume, it can support downstream analyses including tissue segmentation, deformation-based morphometry, and quantitative biomarkers such as regional volumetry or brain-age estimates.

\subsection{Limitations}

{\sloppy
Several limitations warrant discussion. First, extending the framework to multiple MRI sequences will be necessary to better capture treatment- or disease-related abnormalities, such as edema or radiation necrosis, and to make the generated images more interpretable for human readers.

Furthermore, tissue volume changes across different brain regions can have varying impacts on a patient's survival and quality of life, suggesting that accurately predicting changes in smaller yet critical regions may be more informative than capturing changes in larger areas. Global metrics averaged across all voxels may fail to capture such subtle yet clinically significant changes. 

The pretrained autoencoder was developed for CT and contrast-enhanced MRI. Although it reconstructs our pre-contrast T1w data well on visual inspection, the extreme compression ratio can introduce subtle blurring, particularly in areas with fine anatomical detail or pathology, such as resection cavities. This blurriness directly impacts downstream analysis. Intensity-based segmentation algorithms like FSL FAST rely on fine structural detail and distinct intensity gradients to separate tissues. When these gradients are smoothed out by the autoencoder, it leads to widespread misclassifications, most notably between gray matter and CSF around resection cavities, as visible in the error maps (Fig.~\ref{fig:segmentation}). Fine-tuning the autoencoder on non-contrast T1 data could improve these results.

This effect is further exacerbated by the use of T1w-only imaging and the inherent difficulty of segmenting pathological brain tissue. In areas with significant edema, white matter appears darker on T1, causing it to be misclassified as gray matter. Mismatches between the real and predicted scans in these ambiguous regions contribute to artificially lower Dice scores. Indeed, standard intensity-based segmentation algorithms are widely known to be unreliable in the presence of pathological tissue, as normal tissue intensity priors are violated \cite{menzeBRATS2015}. Therefore, these lower Dice scores in pathological regions reflect a limitation of the evaluation method as much as a limitation of the generative model itself.

The SAILOR dataset (25 patients) is small. Although sufficient to demonstrate the approach, it cannot capture the full variability of patient anatomy, tumor location, and treatment response seen in clinical practice. Larger datasets would be required before generalization claims can be made.

Working in latent space adds a step that pixel-space methods avoid: volumes must be encoded before generation and decoded after. This requires precomputing latent embeddings and adds complexity to the pipeline. That said, predictions still only take about 5 seconds on average, with 1.3s for denoising and 3.7s for decoding, showing that the bulk of inference time comes from the decode step itself. For comparison, the 2D baseline takes about 0.3s per slice; processing all 224 slices sequentially would take $\sim$67 seconds.

Finally, we did not model chemotherapy effects. The SAILOR dataset includes temozolomide timing, which the 2D baseline uses as a conditioning variable. Our current 3D model conditions only on dose and time, ignoring systemic therapy, which future work should address.
}

\subsection{Future Directions}

The present work opens several directions for methodological refinement and clinical validation. First, incorporating multiple MRI sequences, particularly T1Gd and T2-FLAIR, would likely improve the prediction of treatment- and disease-related abnormalities such as contrast enhancement, edema, radiation necrosis, and tumor-associated changes. This is also important for clinical interpretation, as follow-up assessment after radiotherapy is not based on T1-weighted imaging alone. Future models should also include chemotherapy and other treatment-related variables to more fully represent the clinical course.

Second, training and validating the framework on larger and more diverse datasets will be essential to assess generalizability across patient populations, tumor locations, imaging protocols, and treatment settings. Future studies should also extend the approach to other intra-axial tumor entities, including IDH-mutant astrocytoma, oligodendroglioma, and brain metastases

Beyond improving image synthesis itself, an important next step is to link predicted anatomical changes to longitudinal clinical outcomes. Integrating cognitive, functional, survival, and quality-of-life measures could help determine whether predicted structural changes translate into clinically meaningful effects. Such outcome-aware evaluation may also clarify which imaging-derived changes are most relevant to patient outcome, rather than merely improving agreement with the observed follow-up image.

Before clinical adoption, the model and its outputs should also be evaluated beyond computational metrics. Reader studies are needed to assess whether clinicians draw consistent conclusions from predicted and real follow-up MRI, and whether the predicted changes provide useful information in cases where treatment-related effects are subtle.

An important future application is counterfactual treatment-plan evaluation. Because the model conditions explicitly on the dose map, one could hold the baseline MRI and target follow-up time fixed while varying clinically acceptable candidate radiotherapy plans. These variations could include, for example, changes in total dose or dose distribution (e.g., focal dose escalation or boosting), fractionation strategy (e.g., hypo- or hyperfractionation), or organ-at-risk sparing policy (e.g., hippocampal avoidance), while still satisfying standard planning requirements for target coverage and clinical acceptability. This would allow alternative treatment strategies to be compared in terms of their predicted anatomical consequences rather than their physical dose properties alone. Such counterfactual virtual follow-up imaging could provide a framework for studying dose-dependent brain injury and normal-brain toxicity, while also supporting patient-specific evaluation of these alternative strategies.

\section{Conclusion}
\label{sec:conclusion}

We present a treatment-aware 3D latent diffusion model for generating post-radiotherapy brain MRI in patients with high-grade glioma. The proposed method incorporates pretreatment T1-weighted MRI, the radiotherapy dose map, and follow-up time as conditioning inputs, allowing for the generation of a full 3D future MRI volume at a given post-treatment time point. By combining latent-space generation with ControlNet-based spatial conditioning, the model directly leverages the patient-specific dose distribution as a driver of predicted anatomical change while maintaining computationally feasible 3D synthesis. This approach has the potential to support virtual follow-up imaging, patient-specific assessment of radiation-induced brain tissue response, and counterfactual evaluation of alternative radiotherapy plans, thereby creating opportunities for more personalized treatment planning.

\section*{Acknowledgment}

The authors would like to thank Oussama Kaddouri for the IT support.

\bibliographystyle{IEEEtran}
\bibliography{references}

\end{document}